\documentclass[aps,prl,preprint,showpacs,groupedaddress]{revtex4}
\usepackage{amsmath}
\usepackage{tipa}
\usepackage{bbm}
\usepackage{txfonts}
\usepackage{graphicx}
\usepackage{dcolumn} 
\usepackage{bm} 
\usepackage{amssymb}
\usepackage{latexsym}

\begin{document}

\preprint{Submitted to PRL}

\title{Test of the universality of free fall with atoms in different spin orientations}

\author{Xiao-Chun Duan}
\author{Xiao-Bing Deng}
\author{Min-Kang Zhou}
\author{Ke Zhang}
\author{Wen-Jie Xu}
\author{Feng Xiong}
\author{Yao-Yao Xu}
\author{Cheng-Gang Shao}
\author{Jun Luo}
\author{Zhong-Kun Hu}\email[E-mail: ]{zkhu@mail.hust.edu.cn}

\affiliation{MOE Key Laboratory of Fundamental Physical Quantities
Measurements, School of Physics, Huazhong University of Science and
Technology, Wuhan 430074, People's Republic of China}

\date{\today}

\begin{abstract}

We report a test of the universality of free fall (UFF) by comparing
the gravity acceleration of the $^{87}$Rb atoms in $m_F=+1$ versus
that in $m_F=-1$, where the corresponding spin orientations are
opposite. A Mach-Zehnder-type atom interferometer is exploited to
sequentially measure the free fall acceleration of the atoms in
these two magnetic sublevels, and the resultant
E$\rm{\ddot{o}}$tv$\rm{\ddot{o}}$s ratio is ${\eta _S}
=(0.2\pm1.2)\times 10^{-7}$. This also gives an upper limit of
$1.1\times 10^{-21}$ GeV/m for possible gradient field of the
spacetime torsion. The interferometer using atoms in $m_F=\pm 1$ is
highly sensitive to the magnetic field inhomogeneity, and a double
differential measurement method is developed to alleviate the
inhomogeneity influence. Moreover, a proof experiment by
modulating the magnetic field is performed, which validates the
alleviation of the inhomogeneity influence in our test. 

\end{abstract}


\maketitle

The universality of free fall (UFF) is one of the fundamental
hypotheses in the foundation of Einstein's general relativity (GR)
\cite{Mis73}. Traditional verifications of the UFF are performed with
macroscopic bodies that weight differently or comprise different
material \cite{Su94,Gun97,Wil04,Sch08,Nie87,Kur89,Car92,Dic94}, achieved a level of 10$^{-13}$
\cite{Gun97,Wil04,Sch08}. There are also lots of work investigating possible
violation of UFF that may be induced by spin-related interactions
\cite{Heh76,Pet78,Mas00,Zha01,Sil07,Ni10,Yas80,Sha02, Hu12}, and UFF tests of this kind have been
performed with polarized or rotating macroscopic bodies
\cite{Hay89,Fal90,Nit90,Win91,Hou01,Zhou02,Luo02,Ni11,Obu14}. In this work two kinds of
proposed spin-related couplings are concerned, namely spin-gravity
coupling and spin-torsion coupling. The corresponding Hamiltonian
operators read as \cite{Hou01,Obu14}
\begin{equation}
\left\{ \begin{array}{l}
{H_{{\rm{spin - gravity}}}} = f(r)\vec S \cdot \hat r\\
{H_{{\rm{spin - torsion}}}} =  - c\vec S_{\rm{Tor}} \cdot \vec S/2
\end{array} \right.,
\label{Eq:1}
\end{equation}
where $\vec S$ is the test mass spin, and $\vec S_{\rm{Tor}}$ stands
for the spacetime torsion. In Eq. (1), $\vec r$ points from the
earth center to the test mass, $f(r)$ is an arbitrary scalar
function of $r$, and $c$ is the light speed. Although UFF tests
involving spin-gravity coupling using polarized macroscopic bodies
have achieved a precision of 10$^{-9}$, the precision decreases
dramatically to a level of 10$^{-5}$ when the result is
re-interpreted in terms of polarized nucleus and even to 10$^{-3}$ in
terms of polarized electron \cite{Ni10,Hou01}. This strongly suggests a
direct UFF test using microscopic test masses to investigate
spin-gravity coupling. On the other hand, for spin-torsion coupling,
it is believed that only matter with intrinsic spin could be
affected by the spacetime torsion \cite{Yas80,Obu14}. And in this sense,
spinful atoms appear to be natural sensors for the torsion
experiments. The spacetime torsion may change along with space, namely
torsion gradient exists, about which there is no information. Here
this information will be explored using spinful quantum particles.

UFF tests with quantum objects have earlier been performed with a
neutron interferometer \cite{Col75}, and in recent years, were carried
out by comparing the free fall acceleration between different atoms
or between atoms and macroscopic masses \cite{Pet99,Fray04,Mer10,Poli11,Bon13,Zhou15,Sch10}. Up
to date, the best precision using quantum objects is $7\times
10^{-9}$ \cite{Pet99}, if the motivations of these tests are not
distinguished. UFF tests on quantum basis are still going on
\cite{Dic13} and tests with higher aimed precision have been
proposed \cite{Agu14,Har15}. As for spin-related UFF tests with quantum
objects, there are few experiments performed. In 2004, the difference
of the free fall acceleration with atoms in two different hyperfine
states has been tested at $1.2\times 10^{-7}$ \cite{Fray04} .
Tarallo et al. \cite{Tar14} performed an UFF test using the
bosonic $^{88}$Sr isotope versus the fermionic $^{87}$Sr isotope at
$1.6\times 10^{-7}$ by Bloch oscillation. In their experiment, the
$^{87}$Sr atoms were in a mixture of different magnetic sublevels,
resulting in an effective sublevel of $\left\langle {{m_F}}
\right\rangle = 0$. They also gave an upper limit for spin-gravity
coupling by analyzing the resonance linewidth broadening caused by possible different
free fall accelerations between different magnetic sublevels.
However, we note that possible anomalous spin-spin couplings
\cite{Win91,Chui93,Gle08} or dipole-dipole interaction \cite{Lah09} between the
$^{87}$Sr atoms with different magnetic sublevels may disturb, or
even cover the spin-gravity coupling effects in their experiment.
Since most models describing spin-related couplings imply a
dependence on the orientation of the spin, we perform a new UFF test
with $^{87}$Rb atoms sequentially prepared in two opposite spin
orientations (Fig. 1), namely $m_F=+1$ versus $m_F=-1$. The
corresponding free fall accelerations are compared by atom
inteferometry \cite{Pet01,Bor02,Mul08,Zhou12,Hu13}, which determines the
spin-orientation related E$\rm{\ddot{o}}$tv$\rm{\ddot{o}}$s ratio
\cite{Eot22} as
\begin{equation}
{\eta _S} \equiv 2({g_ + } - {g_ - })/({g_ + } + {g_ - }),
\label{Eq:2}
\end{equation}
where the gravity acceleration of atoms in $m_F=+1$ ($m_F=-1$) is
denoted as $g_{+}$ ($g_{-}$). This provides a direct way to test
spin-orientation related UFF on quantum basis. And according to Eq.
(1), if the origin of possible violation of UFF is attributed to
spin-torsion coupling, torsion gradient can be linked to $\eta _S$
as
\begin{equation}
{\partial _z}{({S_{{\rm{Tor}}}})_z} =  - \eta _S m ({g_ + } + {g_ -
})/c\Delta {S_z}, \label{Eq:3}
\end{equation}
where $m$ is the atom mass, and $\Delta {S_z}$ stands for the
difference of the spins projection onto vertical direction. Thus
through this kind of UFF test, possible torsion gradient can be also
probed.

\begin{figure}[tbp]
\includegraphics[trim=40 10 40 20,width=0.40\textwidth]{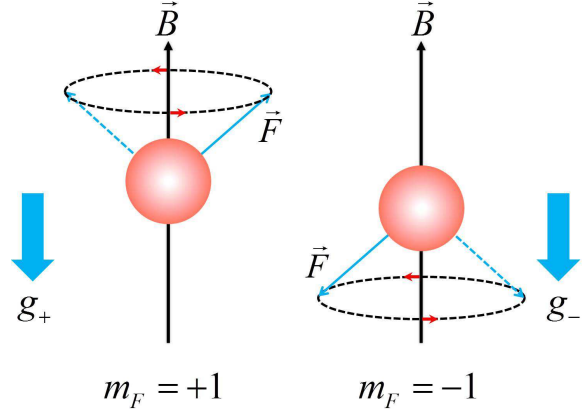}
\caption{\label{fig:1}(color online) Schematic of the spin
orientations for ${}^{87}$Rb atoms in magnetic sublevel $m_F=+1$
versus $m_F=-1$ of the 5$^{2}$S${}_{1/2}$ hyperfine levels. The
bias magnetic field $\vec B$ defines the external direction to which
the atomic spin is referenced. And the total angular momentum $\vec
F$ of each atom processes around $\vec B$.}
\end{figure}

Compared with UFF tests using polarized or rotating macroscopic
masses, it is much easier to prepare atomic ensemble with pure
polarization using stimulated Raman transitions \cite{Mol92}. However,
with atoms in sublevels $m_F=\pm1$, the interferometers are highly
sensitive to the magnetic field inhomogeneity. Thus it is necessary to
select a relatively homogeneous region for interfering. The
magnetic field throughout the shielded interfering tube is mapped
\cite{Zhou10,Hu11}, and the region at about 742 mm height above the
magnetic-optical trap (MOT) center is selected out. The magnetic
field there varies less than 0.1 mG over several millimeters range in
vertical distance with a 115 mG bias magnetic field. Moreover,
compensating coils in anti-Helmholtz configuration are utilized to
further decrease the inhomogeneity. With an injection current of 110
$\rm \mu$A for the compensating coils, the inhomogeneity is decreased by
about one order of magnitude. But the magnetic field shows a binomial variation
along with the vertical distance. The phase shift induced by the magnetic
field inhomogeneity can be calculated by ${\varphi _B} = 2{\alpha}[\int_0^T {B(z(t))dt}  - \int_T^{2T} {B(z(t))dt}]$
\cite{Pet99,Zhou10,Hu11,Dav08},
where $\alpha$ is the strength of first-order
Zeeman shift for $^{87}$Rb atoms in 5$^2{\rm S}_{1/2}$ state, $B(z(t))$
denotes the magnetic field at $z(t)$, and $T$ is the separation time
between Raman laser pulses. Considering a binomial variation model
of $B(z(t)) = B({z_0}) + {\gamma _1}(z(t) - {z_0}) + {\gamma
_2}{(z(t) - {z_0})^2}/2$ (here $\gamma_1$ ($\gamma_2$) is the first
(second) order inhomogeneity coefficient, and $z_0$ stands for an arbitrary
reference point in the selected region), the phase shift induced by the
gravity acceleration and the magnetic field gradient is expressed as
\begin{equation}
\varphi _{{m_F}}^ \pm  =  \mp
{k_{{\rm{eff}}}}{g_{{m_F}}}T_{{\rm{eff}}}^2 + 2{\alpha}{m_F}{T^2}({V_\pi } \mp {V_{\rm r}}/2)[{\gamma _1}
+{\gamma _2}({V_\pi }+gT/4 \mp {V_{\rm r}}/2)T+{\gamma _2}(z_{\rm{s}}-z_0)],\label{Eq:4}
\end{equation}
where the superscript $\pm$  denotes the corresponding direction of
$\emph{\textbf{k}}_{\rm{eff}}$ in the interfering process, with $+$
($-$) indicating the same (opposite) direction between
$\emph{\textbf{k}}_{\rm{eff}}$ and local gravity acceleration. And
${T_{{\rm{eff}}}} \equiv T\sqrt {1 + 2\tau /T + 4\tau /\pi T +
8{\tau ^2}/\pi {T^2}}$ is the effective separation time accounting
for the effect of finite Raman pulses duration ( $\tau$ is the
duration of the $\pi/2$ Raman pulse) \cite{Li15}. In Eq.(4), the
second term corresponds to the effect that induced by the magnetic field
inhomogeneity, where $V_{\rm r}$ is the recoil velocity, $V_{\pi}$ is the
average vertical velocity of the atoms in $F=1$ at the moment of the
interfering $\pi$ pulse (the atoms are initially prepared in $F=1$
before the interfering), and $z_{\rm s}$ is the site where the
interfering process begins.

According to Eq. (4), we take three steps to alleviate the
influence of the magnetic field inhomogeneity in this work. Firstly,
the atomic fountain apex is set near the selected interfering region.
Actually, the time of the interfering $\pi$ pulse is only about 3 ms
near the apex time here. We find this approach is effective to improve our
WEP test. On one hand, with this quasi-symmetrical trajectory for
atoms, the influence of the magnetic field inhomogeneity cancels
significantly. This cancelation assures a relatively long interrogation
time (a separation time as large as $T=25$ ms is allowed here, quite
larger than $T=1$ ms in \cite{Zhou10,Hu11}), which effectively enlarges
the signal of the gravity acceleration. On the other hand, near the
fountain apex, the center of the atomic cloud only moves by a 4.2 mm
vertical distance during the interfering process. In such a small
region, a binomial model for the magnetic field inhomogeneity is appropriate,
which validates the following systematic error correction.
Secondly, as already adopted in typical atom gravimeters \cite{Pet01},
the direction of the effective Raman laser wave vector
$\emph{\textbf{k}}_{\rm{eff}}$ is reversed to make a differential
measurement for each $m_F$. From this reversing
$\emph{\textbf{k}}_{\rm{eff}}$ method, a differential mode
measurement result ($\Delta \varphi _{{m_F}}^{\rm d}\equiv (\Delta
\varphi _{{m_F}}^ +  - \Delta \varphi _{{m_F}}^ - )/2$) and a common
mode measurement result ($\Delta \varphi _{{m_F}}^{\rm c} \equiv (\Delta
\varphi _{{m_F}}^ +  + \Delta \varphi _{{m_F}}^ - )/2$) can be
obtained. There is a residual influence of the magnetic field
inhomogeneity in $\Delta \varphi _{{m_F}}^{\rm d}$ due to the opposite
directions of the recoil velocities between $+k_{{\rm{eff}}}$ and
$-k_{{\rm{eff}}}$ configurations. With only first order magnetic
field inhomogeneity considered, this residual effect can be corrected
using $\gamma_1$ estimated from $\Delta \varphi _{{m_F}}^{\rm c}$. For a
binomial inhomogeneity, more information is required to perform the correction. In
this work, for each $m_F$, a further differential measurement is
performed by modulating $V_{\pi}$ between two values (denoted as
$V_{\pi}^{I}$ and $V_{\pi}^{II}$). We find that the systematic error
correction is simplest by setting $V_{\pi}^{II}=-V_{\pi}^{I}\equiv
-V_{\pi}^0$. Through the double differential measurement, four
combined measurement results can be obtained. The explicit
expressions for $\Delta \Phi _{{m_F}}^{\rm dc} \equiv (\Delta \varphi
_{{m_F}}^{\rm{d}}[V_\pi ^I] + \Delta \varphi _{{m_F}}^{\rm d}[V_\pi ^{II}])/2$
and $\Delta \Phi _{{m_F}}^{\rm cd} \equiv (\Delta \varphi
_{{m_F}}^{\rm c}[V_\pi ^I] - \Delta \varphi _{{m_F}}^{\rm c}[V_\pi ^{II}])/2$
are respectively
\begin{equation}
\left\{ \begin{array}{l} \Delta \Phi _{{m_F}}^{\rm dc} = -
{k_{{\rm{eff}}}}{g_{{m_F}}}T_{{\rm{eff}}}^2 - {\alpha}{m_F}{T^2}{V_{\rm r}}[{\gamma _1}+{\gamma _2}({V_\pi^0 +gT/4 })T]\\
\Delta \Phi _{{m_F}}^{\rm cd} = 2 {\alpha}{m_F}{T^2}{V_\pi^0}[{\gamma _1}+{\gamma _2}({V_\pi^0 +gT/4 })T]
\end{array} \right..\label{Eq:5}
\end{equation}
We note that in the deduction of Eq. (5) from Eq. (4),
$z_s$ varies as $V_{\pi}$. According to Eq. (5), the
residual influence of the magnetic field inhomogeneity in $\Delta
\Phi _{{m_F}}^{\rm dc}$ can be corrected as $\Delta \Phi
_{{m_F}}^{\rm dc}+\Delta \Phi _{{m_F}}^{\rm cd} \times ({V_{\rm r}}/2{V_{\pi}^0
})$, which needs not the knowledge of $\gamma_1$ and
$\gamma_2$. Certainly, with the help of other combined results of
the double differential measurement, $\gamma_1$ and $\gamma_2$ can
be estimated.

In the reversing $\emph{\textbf{k}}_{\rm{eff}}$ differential
measurement, it is important to prepare the atomic ensembles in the
same average velocity between the $+k_{{\rm{eff}}}$ and
$-k_{{\rm{eff}}}$ configurations for each $m_F$, namely $V_{\rm s}^+  =
V_{\rm s}^- $ ($V_{\rm s}$ denotes the average velocity of the atomic ensemble
after the state preparation, and the superscript $\pm$ denotes the
$\emph{\textbf{k}}_{\rm{eff}}$ configuration). Thus the atomic
ensembles experience the same magnetic field inhomogeneity in the
modulation of $\emph{\textbf{k}}_{\rm{eff}}$, ensuring a perfect
differential measurement. Using conventional state preparation method
\cite{Mol92,Lou11}, the equality strongly depends on the pre-determined
Zeeman shift and AC-Stark shift, etc. And the variations of these shifts
will cause opposite changes for $V_{\rm s}^+$ and $V_{\rm s}^- $. Here we
explore an easy but reliable method to guarantee this equality. For
the two interfering configurations, we implement the state
preparations using the Raman lasers both configured in
$+k_{\rm{eff}}$ with the same Raman lasers effective frequency
$\omega _{{\rm{eff}}}$. In this case, for each $m_F$, the state
preparations are completely the same for the two interfering
configurations. Compared with conventional
operation of the interferometer, in addition to usual Raman lasers
frequency chirp, this method needs an extra shift of $\omega_{{\rm{eff}}}$
after the state preparation. This shift will switch
the Raman lasers configuration from $+k_{\rm{eff}}$ to
$-k_{\rm{eff}}$ for the interfering process when needed. As for the modulation of
$V_{\pi}$, the state preparation procedures are totally the same.
And a delay time of about $2V_{\pi}^{0}/g$ ($g\sim9.79$ m/s$^2$) is
inserted in the timing sequence between the state preparation and
the interfering for $V_{\pi}^{II}$, which ensures
$V_{\pi}^{II}=-V_{\pi}^{I}$.

The experiment is performed in an atom gravimeter detailedly
reported in Ref. \cite{Zhou12}. It takes 750 ms to load about 10$^8$
cold $^{87}$Rb atoms from a dispenser using a typical MOT. Then the
atoms are launched upward and further cooled to about 7 ${\rm \mu}$K with
a moving molasses procedure in the atomic fountain. After a flight
time of 356 ms from the launch, a Raman $\pi$ pulse with a duration
of 46 $\rm \mu$s is employed to implement the state preparation. Then
the unwanted atoms are removed by a blow-away beam. When arriving
at 742 mm height, the atomic cloud undergoes the $\pi/2-\pi-\pi/2$
Raman pulses with a pulse separation time of $T=25$ ms. The
transition probability of the atoms after the interfering is
obtained through normalized fluorescence detection. The entire
process of a single shot measurement as described above takes 1.5 s.
Before the formal data acquisition, $V_{\pi}^{I,II}$  should be
measured for each $m_F$ to calculate the correction. This velocity
is obtained from the spectroscopy of the VSRT \cite{Mol92} with a Raman
$\pi$ pulse applied at the right moment. The measured average
velocities are $V_{\pi}^{I}=30.6(1)$ mm/s and
$V_{\pi}^{II}=-30.6(1)$ mm/s for the selected atoms in both $\left| {F =
1,{m_F} = + 1} \right\rangle$ and  $\left| {F = 1,{m_F} = - 1} \right\rangle$
(it takes about 180 s to measure one average velocity. In reality, these velocities will change in long term, which will be discussed later).

Finally, the measurement of the gravity acceleration of the atoms in
different magnetic sublevels is performed sequentially. For each $m_F$, one full
interferometry fringe is obtained by scanning the chirp rate of
$\omega _{{\rm{eff}}}$  in 20 steps in each $\emph{\textbf{k}}_{\rm{eff}}$
configuration for each $V_{\pi}$, namely 30 s for a full fringe. Meanwhile, in order to
reduce the effect of possible long-term drift, eight adjacent
fringes are grouped as a cycle unit, with one fringe corresponding
to one combination of $\emph{\textbf{k}}_{\rm{eff}}$, $V_{\pi}$ and $m_F$.
The switches between the combinations are automatically
controlled by the computer.
It takes 10 hours to repeat the cycle unit for 150 times, and
the phase shifts are extracted by the cosine fitting from the
fringes. The Allan deviation calculated from the consecutive
measurement of $\Delta \Phi _{{m_F}}^{\rm dc}$ shows a short-term
sensitivity of about 3.5$\times 10^{-6} g$/$\sqrt{\rm{Hz}}$ for the
gravity acceleration measurement with each $m_F$.

\begin{figure}[tbp]
\includegraphics[trim=40 10 40 20,width=0.40\textwidth]{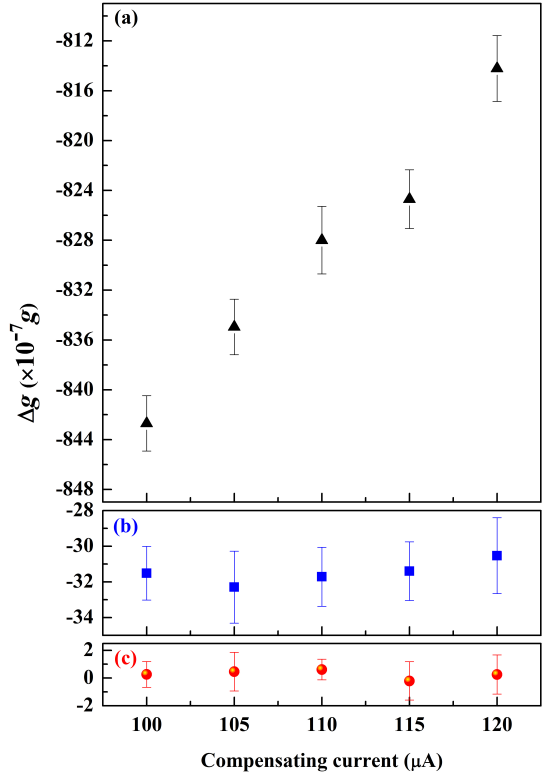}
\caption{\label{fig:2}(color online) Estimated $\Delta g$ using
different methods in the modulation of the injection current for the
compensating coils. The data acquiring time for each injection current is only one hour.
In figure (a) no measures are taken to decrease
the magnetic field inhomogeneity influence; In figure (b) the conventional
differential measurement is adopted to decrease the influence, and
in figure (c) the influence is alleviated using the double
differential measurement. We note that the error bars in these
figures only refer to the statistics standard deviation.}
\end{figure}

In order to validate the efficiency of alleviating the influence of
the magnetic field inhomogeneity in our WEP test by this double
differential measurement, in addition to the 110 $\rm \mu$A injection
current for the compensating coils, tests with other four values of
the current are also performed. And the result is shown in Fig. 2,
which is reported as $\Delta g \equiv {g_ + } - {g_ - }$ for each
injection current (the error bars are only the corresponding statistical
standard deviations). In Fig. 2(a), $\Delta g$ is estimated by
$\Delta g = (\Delta \varphi _{{m_F} =  + 1}^ +  - \Delta \varphi
_{{m_F} = - 1}^ + )/{k_{\rm{eff}}}T_{\rm{eff}}^2$, namely the situation without any differential
measurement to decrease the inhomogeneity influence. In Fig. 2(b),
$\Delta g$ is estimated by
\begin{equation}
\Delta g = \frac{1}{{{k_{\rm{eff}}}T_{\rm{eff}}^2}} [(\Delta \varphi _{{m_F} =  +
1}^{\rm d} + \frac{{{V_{\rm r}}}}{{2V_\pi ^0}}\Delta \varphi _{{m_F} =  +
1}^{\rm c} )- (\Delta \varphi _{{m_F} =  - 1}^{\rm d}+\frac{{{V_{\rm r}}}}{{2V_\pi ^0}}\Delta \varphi _{{m_F} =  - 1}^{\rm c})],\label{Eq:6}
\end{equation}
which is only capable of eliminating the effect
of the first order magnetic field inhomogeneity.
And in Fig. 2(c), $\Delta g$ is estimated by
\begin{equation}
\Delta g = \frac{1}{{{k_{\rm{eff}}}T_{\rm{eff}}^2}}[(\Delta \Phi _{{m_F} =  +
1}^{\rm dc} + \frac{{{V_{\rm r}}}}{{2V_\pi ^0}}\Delta \Phi _{{m_F} =  +
1}^{\rm cd}) - (\Delta \Phi _{{m_F} =  - 1}^{\rm dc} +
\frac{{{V_{\rm r}}}}{{2V_\pi ^0}}\Delta \Phi _{{m_F} =  -
1}^{\rm cd})].\label{Eq:7}
\end{equation}
According to Fig. 2(a) ,  the influence of the
magnetic field inhomogeneity changes dominantly with the injected
current. In Fig. 2(b), the influence of the inhomogeneity has been suppressed by about a factor of 26,
but there is still a considerable residual effect. In Fig. 2(c), there is no obvious dependence of $\Delta g$ on
the injection current. And what is more important, the residual effect is also suppressed below the level of $10^{-7}g$, which proofs that our correction based on this
double differential measurement is quite effective. The statistics result
for the 110 $\rm \mu$A compensating current with the 10 hours measurement
time is $(-1\pm 3) \times 10^{-8}g$ without any other corrections. It also shows
that a majority of the phase shift due to the magnetic field
inhomogeneity is canceled in the double differential measurement.
The residual effect due to the Raman pulses durations is thus safely
neglected here.

In this differential measurement of the gravity acceleration, some
disturbances, for example, that are induced by the AC-Stark shift, can
be largely suppressed, and other disturbances, for example, that
are induced by nearby masses or tilt of the Raman lasers, are common for
the atoms in $m_F=+1$ and $m_F=-1$ and are cancelled in the final
comparison. The main systematic error still comes from the effect
associated with the magnetic field inhomogeneity. In this work the
equality of $V_{\rm s}^+$ and $V_{\rm s}^-$ is well guaranteed  by our special
state preparation. However the value of $V_{\rm s}^+$ and $V_{\rm s}^-$ drifts
in a common way due to the change of the AC-Stark shift, which is
induced by the variation of the Raman laser power as the room
temperature changes periodically. A peak-to-peak variation of 0.25 mm/s
for $V_{\pi}$ is observed, which on one hand affects the cancelation
in the double differential measurement and on the other hand limits the
accuracy of the correction $\Delta \Phi _{{m_F}}^{\rm dc}+\Delta \Phi
_{{m_F}}^{\rm cd} \times ({V_{\rm r}}/2{V_{\pi}^0 })$. The corresponding
contributed uncertainty on $\Delta g$ is $1.2\times 10^{-7}g$.
We notice that there is a
difference of $2\pi \times 2.8$ kHz/G for $\alpha$ between
$^{87}$Rb atoms in $F=1$ versus $F=2$, and the corresponding error
is about $-3 \times 10^{-8}g$. Though the magnetic field
inhomogeneity in the selected interfering region shows a binomial
variation, higher order inhomogeneity has been investigated as well. We extent the
calculation of Eq. (4) to the case of a third order magnetic field
inhomogeneity present, and find our double differential measurement
capable to alleviate the influence of the third order inhomogeneity
by amount of 70 percent. The final resultant
E$\rm{\ddot{o}}$tv$\rm{\ddot{o}}$s ratio (calculated by $\Delta
g/g$) is $(0.2\pm 1.2)\times 10^{-7}$, which indicates that the
violation of UFF has not been observed at the level of $1.2\times
10^{-7}$ for the atoms with different polarization orientations.
According to Eq. (3), this corresponds to a constrain of $1.1\times
10^{-21}$ GeV/m for possible gradient field of spacetime torsion
(for this experiment, $\Delta {S_z}$ is 2$\left| {{m_F}} \right|
\hbar$). We note that the bias magnetic field direction is crucial
in the deduction of this constrain, since it defines the reference of
the spin orientation.

In conclusion, we have tested UFF with atoms in different spin
orientations based on a Mach-Zehnder-type atom interferometer, and
the violation of UFF is not observed at the level of $1.2\times
10^{-7}$. This work represents the first direct spin-orientation
related UFF test on quantum basis, and possible spacetime torsion
gradient is also constrained to an upper limit of $1.1\times
10^{-21}$ GeV/m. We anticipate that the precision of this kind of
UFF test will be improved by constructing a more homogeneous
magnetic field or by exploiting internal-state invariant atom
interferometers (see Ref. \cite{Fray04,Dav08,Lev09,Berg15,Alt13},
for example). In this work, in order to achieve this precision, on
one hand, the fountain apex is set near the selected interfering
region, and on the other hand, the double differential measurement
method as well as the special state preparation method is developed. These
strategies may be illuminating for other high precision measurements
using atom interferometry.

We thank Yuanzhong Zhang for discussions about the UFF background,
Weitou Ni for discussions about UFF tests with macroscopic polarized
bodies, Xiangsong Chen, Jianwei Cui and Yungui Gong for discussions about
spin-torsion coupling, and Zhifang Xu for discussions about
dipole-dipole interaction between atoms. This work is supported by
the National Natural Science Foundation of China (Grants Nos.
41127002, 11574099, and 11474115) and the National Basic Research
Program of China (Grant No. 2010CB832806).


\end{document}